\documentclass{aa}

\usepackage{graphicx}
\usepackage{txfonts}
\usepackage{hyperref}

\hypersetup{
  colorlinks=true,   %% links colored instead of frames
  urlcolor=blue,     %% external hyperlinks
  linkcolor=blue,     %% internal latex links (eg Fig)
  citecolor=blue
}

\usepackage{natbib}
\bibpunct{(}{)}{;}{a}{}{,}
\usepackage{url}
\let\textcite\citet
\let\cite\citep 

\begin{document}

   \title{A discontinuity in the luminosity-mass relation and fluctuations in the evolutionary tracks of low-mass and low-metallicity stars at the Gaia M-dwarf gap}

   \author{Santana Mansfield
          \inst{1}\fnmsep\thanks{Email: smansfield@astro.uni-bonn.de}
          \and
          Pavel Kroupa\inst{1,2}
          }

   \institute{Helmholtz-Institut f\"ur Strahlen- und Kernphysik (HISKP), Universit\"at Bonn, Nu\ss allee 14-16, D-53115 Bonn, Germany
         \and
             Astronomical Institute, Faculty of Mathematics and Physics, Charles University in Prague, V  Hole\v{s}ovi\v{c}k\'ach 2, CZ-18000 Praha, Czech Republic\\
             }

   \date{Received 11 February 2021 / Accepted 23 April 2021}

% \abstract{}{}{}{}{} 
% 5 {} token are mandatory
 
  \abstract
  % context heading (optional)
   {The Gaia M-dwarf gap is a recently discovered feature in the colour-magnitude diagram that shows a deficiency of low-mass and low-metallicity stars at the lower end of the main sequence.}
  % aims heading (mandatory)
   {We aim at performing theoretical stellar modelling at low metallicities using a fine mass step and a fine time step, looking specifically for the transition of models from partially to fully convective, since the convective kissing instability that occurs at this transition is believed to be the cause of the gap.}
  % methods heading (mandatory)
   {Stellar evolution models with metallicities of Z = 0.01, Z = 0.001 and Z = 0.0001 are performed using MESA, with a mass step of 0.00025 M$_{\odot}$ and a time step of 50,000 years.}
  % results heading (mandatory)
   {The small time step produced models that experience loops in their evolutionary tracks in the Hertzsprung-Russell (HR) diagram. The fluctuations in effective temperature and luminosity correspond to repeated events in which the bottom of the convective envelope merges with the top of the convective core, transporting $^3$He from the core to the surface. In addition to the episodes of switching from partially to fully convective, several near-merger events that produced low amplitude fluctuations were also found. Low-metallicity models undergo the convective kissing instability for longer portions of their lifetime and with higher fluctuation amplitudes than models with higher metallicities. The small mass step used in the models revealed a discontinuity in the luminosity-mass relation at all three metallicities.}
  % conclusions heading (optional), leave it empty if necessary 
   {The repeated merging of the convective core and envelope, along with several near-merger events, removes an abundance of $^3$He from the core and temporarily reduces nuclear burning. This results in fluctuations in the star's luminosity and effective temperature, causing loops in the evolutionary track in the HR diagram and leading to the deficiency of stars at the M-dwarf gap, as well as a discontinuity in the luminosity-mass relation.}

               
    \titlerunning{Modelling the M-dwarf gap}
    \authorrunning{S. Mansfield \& P. Kroupa}

   \maketitle
%
%________________________________________________________________

\section{Introduction}

\vspace{4.7pt}
Stars are formed from the gravitational infall of material in molecular clouds, which produces many stars with masses lower than the mass of the Sun. A substantially greater amount of material is needed for stars with masses higher than the Sun, and so these stars are formed in much fewer numbers. Stars undergo nuclear reactions in their cores for masses $m \gtrsim 0.08\ \textnormal{M}_{\odot}$, and thus the majority of all stars have masses from this lower limit up to the solar mass. 

The stellar luminosity function and the stellar mass function are used to make comparisons across all types of stars, and these functions depend sensitively on the luminosity-mass relation. This relation has long been considered a smooth and differentiable function that exhibits structure correlating to the features of the stellar luminosity function \cite{pavel}. For example, an inflection in the luminosity-mass relation at $\approx 0.30\ \textnormal{M}_{\odot}$ was first discussed by \textcite{copeland} as the H$_2$ dissociation zone having a lower adiabatic gradient. The minimum in the first derivative of the luminosity-mass relation, $dm/dM_{bol}$, at this mass was found to correspond to the maximum of the observed luminosity function of low-mass stars \cite{pavel, pavelmasslum} due to H$_2$ formation as well as stars becoming fully convective at this mass. Thus, all mono-age and mono-metallicity stellar populations show a pronounced and sharp maximum in the stellar luminosity function at a visual absolute magnitude of $M_{\rm V}\approx 12$ \cite{Kroupa02, Kroupa+13}.
The luminosity-mass relation has long been known to be metallicity dependant \cite{elson, pavelmasslum}.

A recently discovered and related feature, called the Gaia M-dwarf gap, or Jao gap, was found in the observational \textit{Gaia Data Release 2 (DR2)} data \cite{jao} and coincides with the same stellar mass of the inflection in the luminosity-mass relation at $\approx 0.30\ \textnormal{M}_{\odot}$. This feature lies at the lower end of the main sequence (MS) in the G$_{BP} - $G$_{RP}$, M$_{G}$ colour-magnitude diagram (CMD) and represents a deficiency in the density of stars (and subsequently a dip in the luminosity function). This gap was recently confirmed in the new \textit{Gaia Early Data Release 3 (EDR3)} data \cite{dr3} at Gaia magnitudes $2.2 < \textnormal{G}_{BP} - \textnormal{G}_{RP} < 2.8$ and 10.0 $<$ M$_G <$ 10.3 and portrays a drop in density by 17 $\pm$ 6\% \cite{jao}. It is postulated that this occurs due to M-dwarf stars with masses around $\approx 0.30 - 0.35\ \textnormal{M}_{\odot}$ undergoing the transition from being partially to fully convective (\citealp{jao}; \citealp{baraffe}).

The thermonuclear process for low-mass stars is governed by the proton-proton I (ppI) chain:

\vspace*{-2pt}
\begin{equation}
\ \ \ \hspace*{60pt}    p + p \longrightarrow d + e^+ + \nu_e \ \ \ \  \label{eq1}
\end{equation}

\vspace*{-21.5pt}
\begin{equation}
\ \ \ \hspace*{60pt}      p + d \longrightarrow\ ^3\!He + \gamma \label{eq2}
\end{equation}

\vspace*{-21.5pt}
\begin{equation}
\ \ \ \hspace*{60pt}      ^3He +\ ^3\!He \longrightarrow\ ^4\!He + 2p  \label{eq3} 
\end{equation}

\vspace{4.75pt}
\noindent where two protons ($p$) combine to make deuterium ($d$) by releasing a positron ($e^+$) and an electron neutrino ($v_e$), and ultimately synthesise a $^4$He nucleus with the release of a photon ($\gamma$) and via two $^3$He nuclei. Equation \ref{eq3} becomes important once the central temperature is $\textnormal{T} \gtrsim 7 \times 10^6$ K \cite{dearborn}, that is, $m \gtrsim 0.26\ \textnormal{M}_{\odot}$. Without this reaction, the energy produced by the other two reactions (Eqs. \ref{eq1} and \ref{eq2}) is not great enough for the material in the core to become unstable to convection \cite{baraffe}, which occurs when the radiative temperature gradient is larger than the adiabatic gradient, $\nabla_{rad} > \nabla_{ad}$, where

\begin{equation}
\hspace*{72pt}
    \nabla_{rad} \propto L\kappa / T^4
    \label{e4}
\end{equation}

\vspace{5.5pt}
\noindent When the reaction given by Eq. (\ref{eq2}) dominates, the $^3$He abundance grows, and at a temperature $\textnormal{T} \gtrsim 7 \times 10^6$ K, Eq. (\ref{eq3}) produces enough energy in the core for the centre to become convective \cite{chabrier1997}. Then, for example, as a $0.30\ \textnormal{M}_{\odot}$ star approaches the MS, it has a radiative core, but the temperature increase due to the pre-MS contraction means that the production and destruction of $^3$He by Eqs. (\ref{eq2}) and (\ref{eq3}) releases enough energy for the material in the core to become unstable to convection; as such, the star reaches the MS with a convective core and envelope, which are separated by a thin radiative layer \cite{ezer}.

Before the discovery of the M-dwarf gap, \textcite{vansaders} theorised that at the mass directly above the convective boundary, stars undergo the `convective kissing instability'. As $^3$He is destroyed at a slower rate by Eq. (\ref{eq3}) than is produced by Eq. (\ref{eq2}) at these low-mass temperatures \cite{baraffe}, it cannot reach equilibrium abundance. The abundance of $^3$He increases, and the core grows in size until it comes into contact with the convective envelope, resulting in periods of full convection \cite{vansaders}. The mixing during these periods carries $^3$He out towards the surface, nuclear reactions in the core subside, and the core contracts and becomes separated again from the envelope \cite{vansaders}. This continues periodically, with variations in the star's luminosity and effective temperature, until the abundance of $^3$He is high enough throughout the star that it remains fully convective for the remainder of its lifetime. The variations from this convective kissing instability are thought to result in the M-dwarf gap in the density of observed stars seen at these temperatures and luminosities. \textcite{baraffe} produced models in steps of $0.01\ \textnormal{M}_{\odot}$ and note the merging of the core and envelope in their $0.34\ \textnormal{M}_{\odot}$ and $0.36\ \textnormal{M}_{\odot}$ models, along with a decrease in central $^3$He abundance and a change in the slope of the luminosity-mass relation at these masses. \textcite{macdonald} also find this relation and a dip in the luminosity function for their $0.33\ $--$\ 0.35\ \textnormal{M}_{\odot}$ models. \textcite{feiden} reproduced the M-dwarf gap in a CMD made from population synthesis models and also find periodic pulsations in luminosity, radius, and core temperature due to the $^3$He instability. 

The Gaia M-dwarf gap is prominent across the blueward edge of the MS, where lower-metallicity stars are found.
\textcite{feiden} show that their models with low metallicity ([Fe/H]$\ =\ $--$\ 0.7$) encounter this instability at lower masses (and thus cooler core temperatures) than models with higher metallicities; from this they conclude that the dependence of temperature on stellar mass is one of the critical factors for this instability. At low temperatures, the opacity dependence of the radiative temperature gradient (Eq. \ref{e4}) becomes important. For lower metallicity, the opacity is decreased and the central temperature needed for the energy transport in the core to occur by radiative transport is lower \cite{chabrier1997} and thus requires less mass. As such, the convective kissing instability is expected to happen at lower masses for lower metallicities since the higher-mass models have the temperature necessary for radiative cores. This work aims to investigate the dependence of the M-dwarf gap on the very low metallicities found in old globular clusters. Based on this, we suggest that the M-dwarf gap should be well represented in metal-poor globular cluster CMDs, although the low stellar masses at which this effect is found will be difficult to detect in clusters at great distances. 

This work uses stellar evolution models to reproduce the convection-mass transition and investigate a fine grid of masses and very low metallicities with a short-scale time step. These models are described in Sect. \ref{stellar models}, the results are presented in Sect. \ref{results}, and we conclude with a discussion in Sect. \ref{discussion}.

\vspace*{8.5pt}
\section{Stellar models}\label{stellar models}

\vspace*{10.7pt}
The 1D stellar evolution code Modules for Experiments in Stellar Astrophysics
(MESA, version 15140) was used (\citealp{Paxton2011}, \citealp{Paxton2013}, \citealp{Paxton2015}, \citealp{Paxton2018}, \citealp{Paxton2019}). The equations of state (EOSs) used in MESA are a combination of the SCVH \cite{scvh}, OPAL \cite{opal}, HELM \cite{helm}, FreeEOS \cite{Irwin2004}, and PC \cite{Potekhin2010} EOSs. The opacity tables relevant to this work are given by \textcite{gs}, and OPAL (\citealp{Iglesias1993},
 \citealp{Iglesias1996}); with additionally the electron conduction opacities from
  \textcite{Cassisi2007} and the opacity tables for the low temperatures of M-dwarf atmospheres in which molecules and dust grains are able to form are from \textcite{ferguson}. For the outer boundary conditions we adopted the prescriptions provided by the \textcite{hauschildt} model atmosphere tables, taken at $\tau = 100$ \cite{Paxton2011, chabrier1997}. The nuclear reaction rates are from \textcite{Cyburt2010}, and we included additional weak reaction rates \cite{Fuller1985, Oda1994, Langanke2000}. Screening was included via the prescription of \textcite{Chugunov2007}. Thermal neutrino loss rates are from \textcite{Itoh1996}. 
 The initial helium abundance is Y = 0.24 + 2Z by mass fraction. Material becomes unstable to convection under the Schwarzschild criterion, and convective energy transport is described using mixing length theory. 
  
  \begin{figure}
    \centering
    \includegraphics[width=\columnwidth]{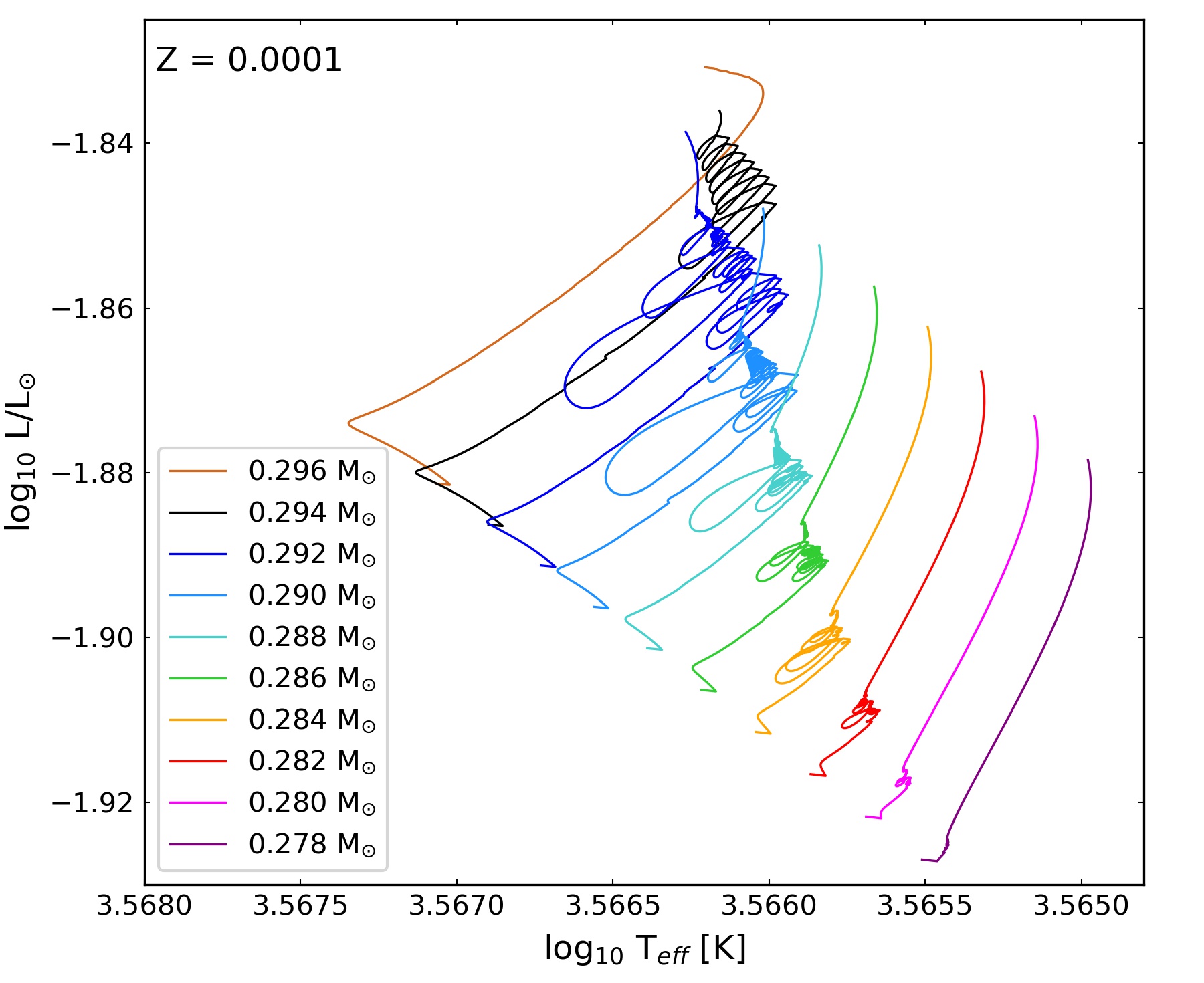}
    \includegraphics[width=\columnwidth]{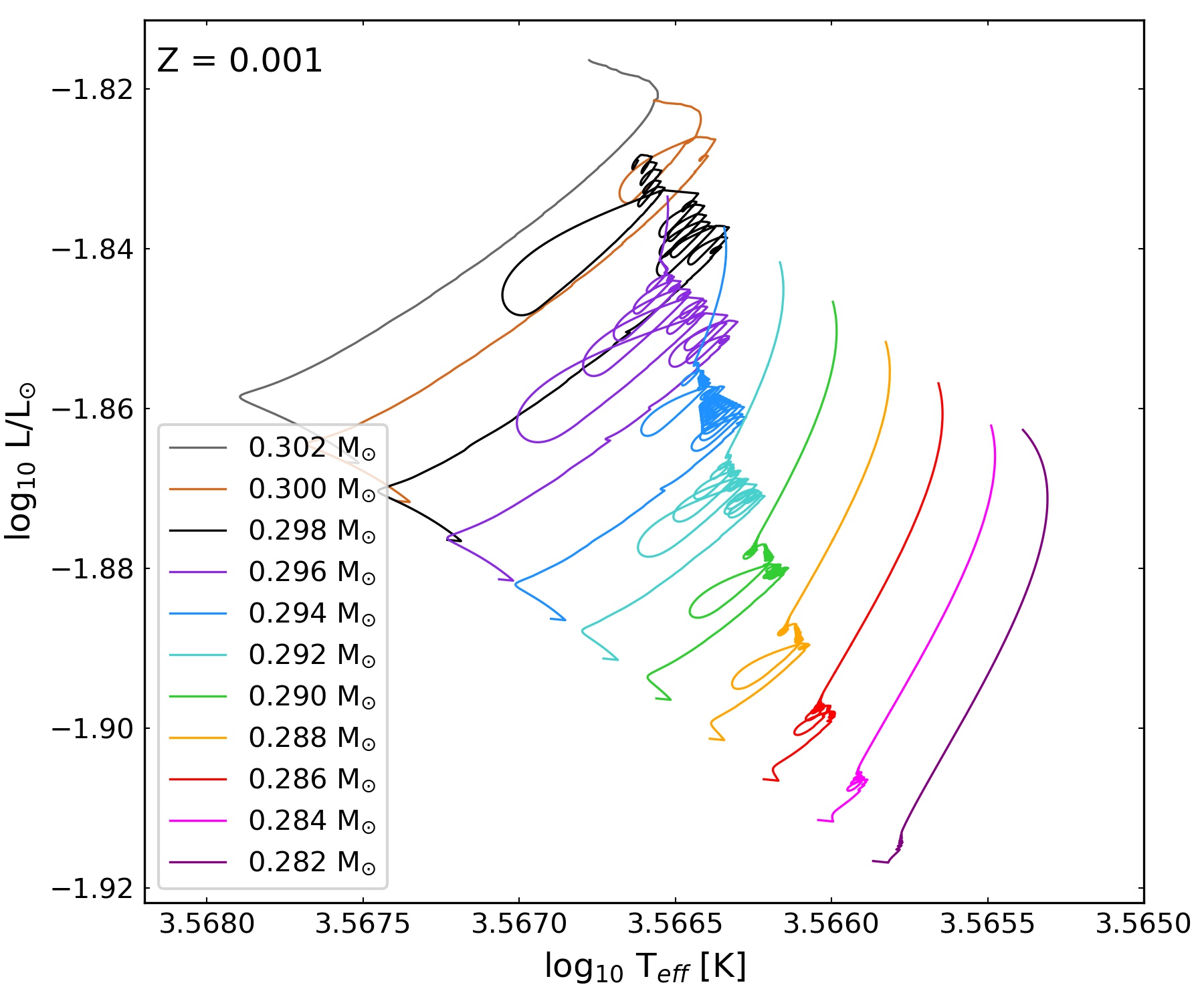}
    \includegraphics[width=\columnwidth]{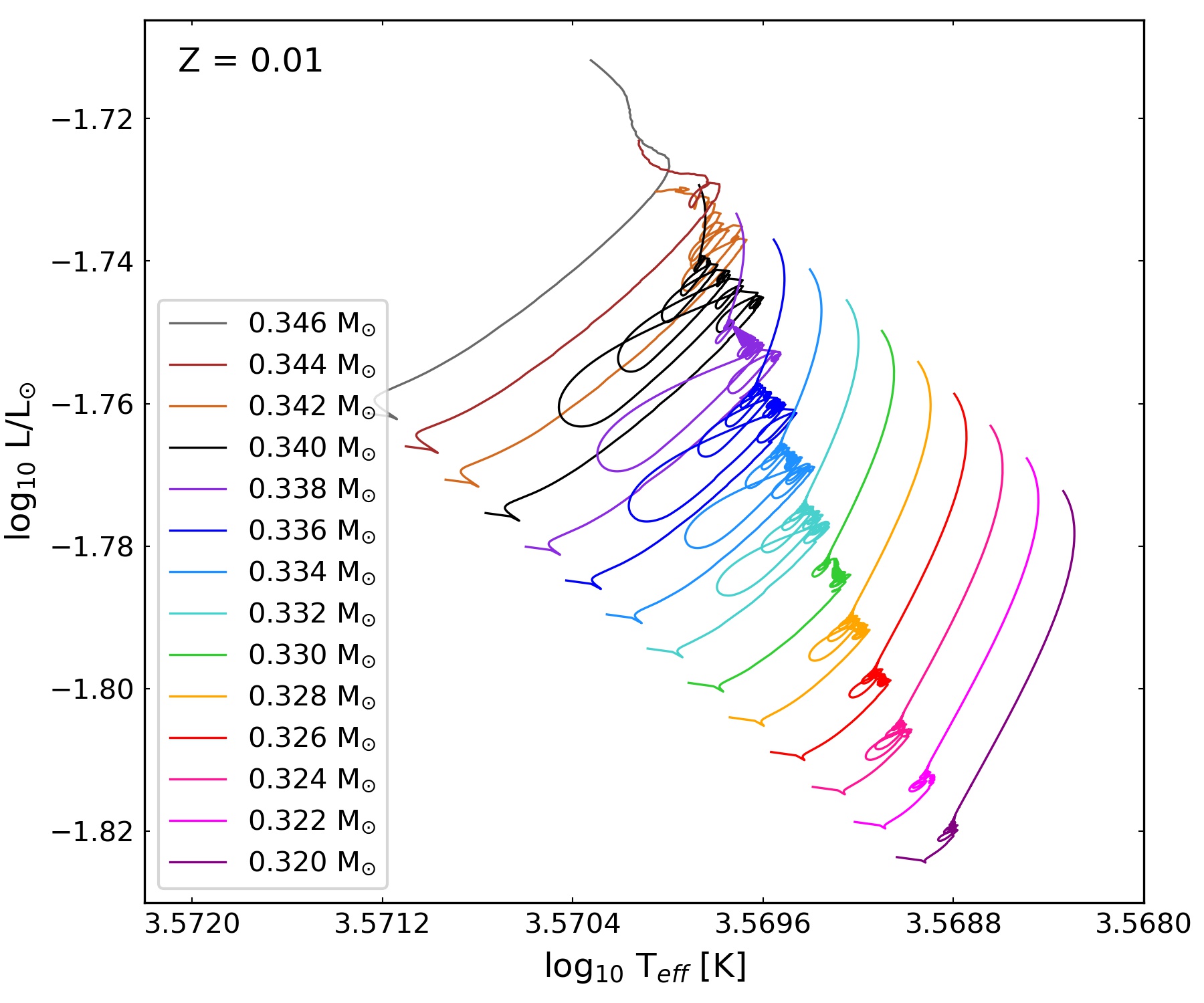}
    
    \vspace*{8pt}
    \caption{Evolutionary tracks shown in the HR diagram for three sets of stellar models with metallicities Z = 0.0001 (top), Z = 0.001 (middle), and Z = 0.01 (bottom). The models begin their evolution on the left and move upwards along the tracks to reach the final points at $10$ Gyr. The stars are on the MS throughout this evolution.}
    \label{HR_all}
\end{figure}

  Three initial metallicities were chosen (Z = 0.01, Z = 0.001, and Z = 0.0001) in order to investigate the M-dwarf gap on the blue edge of the MS and investigate the low metallicities found in globular clusters. An initial set of models for each metallicity was computed using a mass step of 0.01 M$_{\odot}$ to determine at which mass the models begin to have radiative cores. Once found, model sets were then produced using a smaller mass step of 0.00025 M$_{\odot}$. The models were calculated from the zero-age main sequence (ZAMS) up to 10 Gyr with a maximum time step of 50,000 years. One final model with an initial mass of $0.28\ \textnormal{M}_{\odot}$ and Z = 0.0001 was computed up to 1 Gyr using a maximum time step of 1,000 years.

  \begin{figure*}
    \centering
    \hspace*{-8pt}
    \includegraphics[scale=0.292]{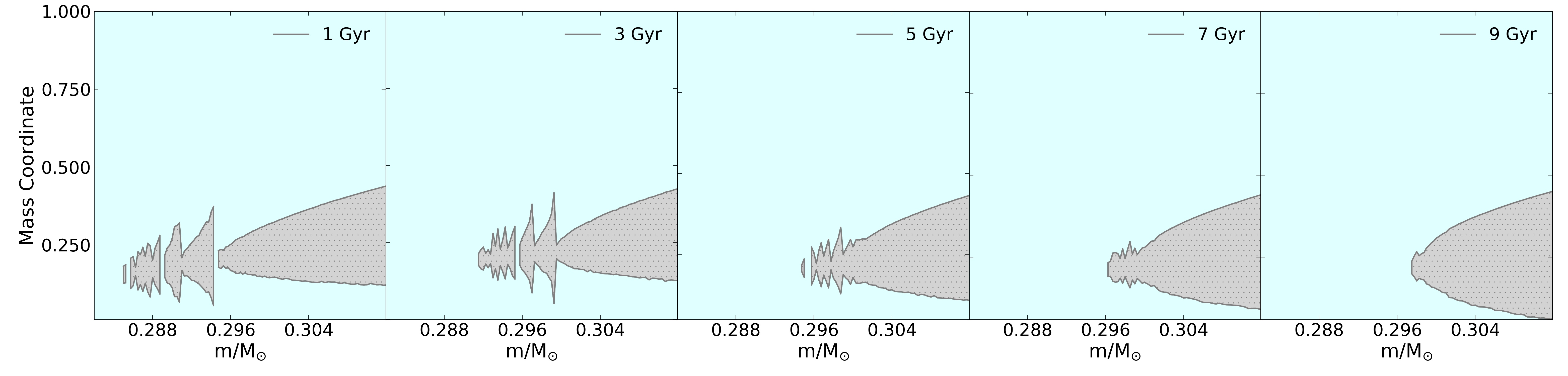}
    \caption{Mass coordinates of the convective (blue) and radiative (grey dotted) regions within the Z = 0.001 models over 1, 3, 5, 7, and 9 Gyr. The models that experience a merger of the convective core and envelope can be seen as gaps in the radiative region. A movie of the full time lapse is available online.}
    \label{mass_rad}
\end{figure*}

\begin{figure}
    \centering
    
    \vspace*{-4pt}
    \includegraphics[width=\columnwidth]{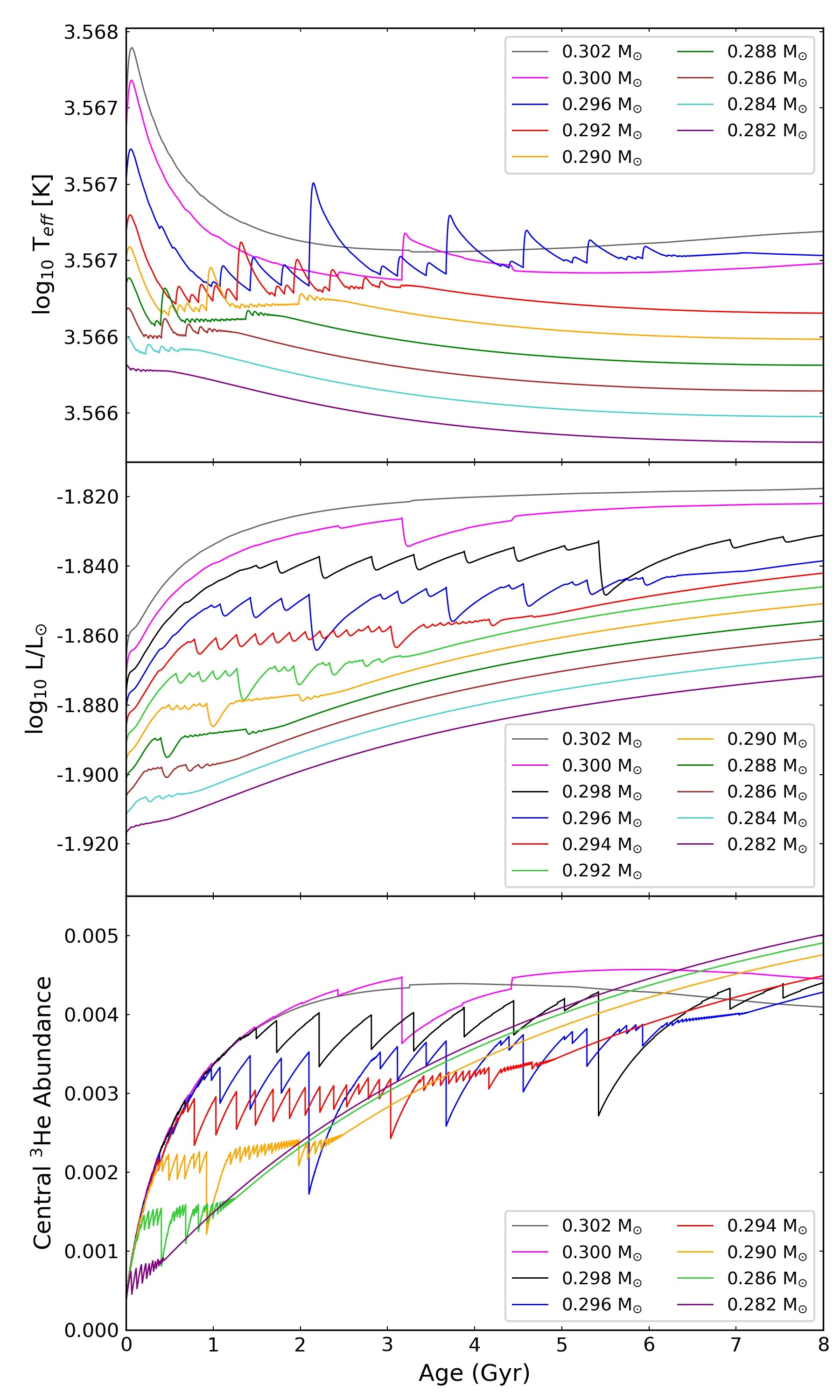}
    
    \vspace*{-2pt}
    \caption{Temperature (top) luminosity (middle), and central $^3$He abundance (by mass fraction; bottom) over time for models with a metallicity of Z = 0.001. The fluctuations occur for several billion years for the higher-mass models.}
    \label{lumtempage}
\end{figure}

\vspace{4pt}
\section{Results}\label{results}

\vspace{9pt}
The low-mass MS models slowly undergo nuclear burning via the ppI chain described by Eqs. (\ref{eq1}) - (\ref{eq3}). By the time they reach an age of 10 Gyr, the hydrogen abundance in the core has only just begun to deplete. The Hertzsprung-Russell (HR) diagrams in Fig. \ref{HR_all} (and also later in Fig. \ref{convmassage}) show the models experiencing a period of fluctuations that appear as loops along their evolutionary tracks during the MS. The age of the model at which this occurs increases with increasing mass. The mass range for this instability occurs at a higher mass for a higher metallicity, and the range of masses is larger for the higher-metallicity models. For the masses below the convective boundary, the models are fully convective throughout their lifetimes. Between a certain mass range for each metallicity (given in Table \ref{massranges}), the models begin their lives with a convective core, a thin radiative layer, and a convective envelope. They then undergo a period whereby they repeatedly switch from this structure to one of full convection and then relax into a fully convective state for the remainder of their lifetimes. This is illustrated in Fig. \ref{mass_rad}, which shows the mass coordinates of the convective and radiative regions of the Z = 0.001 models over time. At 1 Gyr, the models with mass $m < 0.28600$ M$_{\odot}$ are fully convective, models with mass $0.28600$ M$_{\odot} \leq m \geq 0.29845$ M$_{\odot}$ have begun the convective kissing instability, and the models with higher masses have a stable convective core and radiative region. The 0.28550 M$_{\odot}$, 0.28900 M$_{\odot}$, and 0.29450 M$_{\odot}$ models are also fully convective at 1 Gyr, having undergone a merger of the convective core and envelope at this time. At 3 Gyr, models with mass $m < 29275\ \textnormal{M}_{\odot}$ and the 0.29550 M$_{\odot}$ model that has merged are fully convective. At 5 Gyr, the models with mass $m < 0.29500$ M$_{\odot}$ along with the 0.29525 M$_{\odot}$ and 0.29550 M$_{\odot}$ models that have merged at this time are fully convective. By 9 Gyr, models with mass $m < 0.29775\ \textnormal{M}_{\odot}$ are fully convective, and models with mass $m\ >\ 0.31150\ \textnormal{M}_{\odot}$ have a radiative core.

\vspace{5pt}
\subsection{Convective kissing instability}

\begin{figure}
     \centering
     \includegraphics[width=\columnwidth]{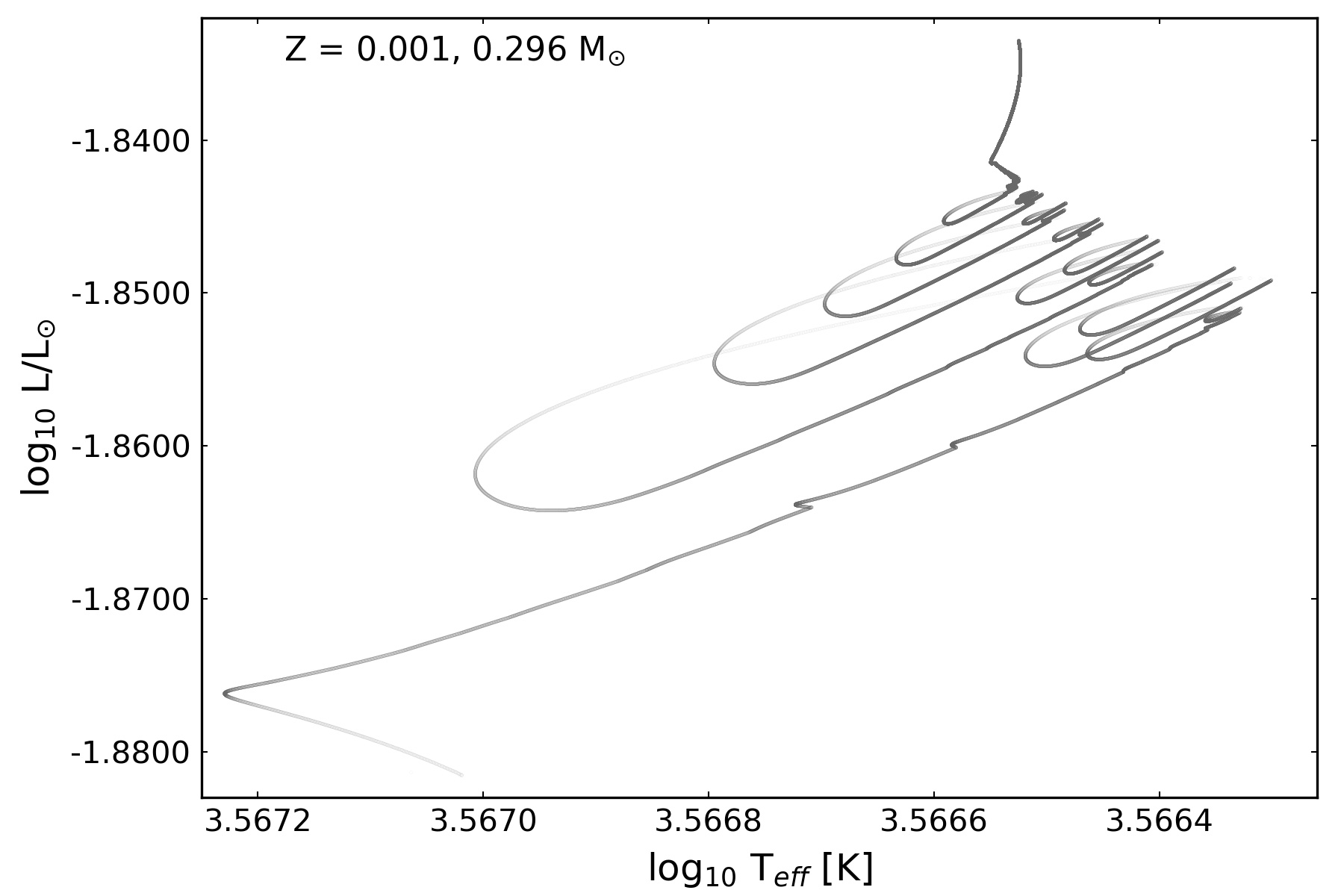}
     
     \vspace*{-1pt}
     \includegraphics[width=\columnwidth]{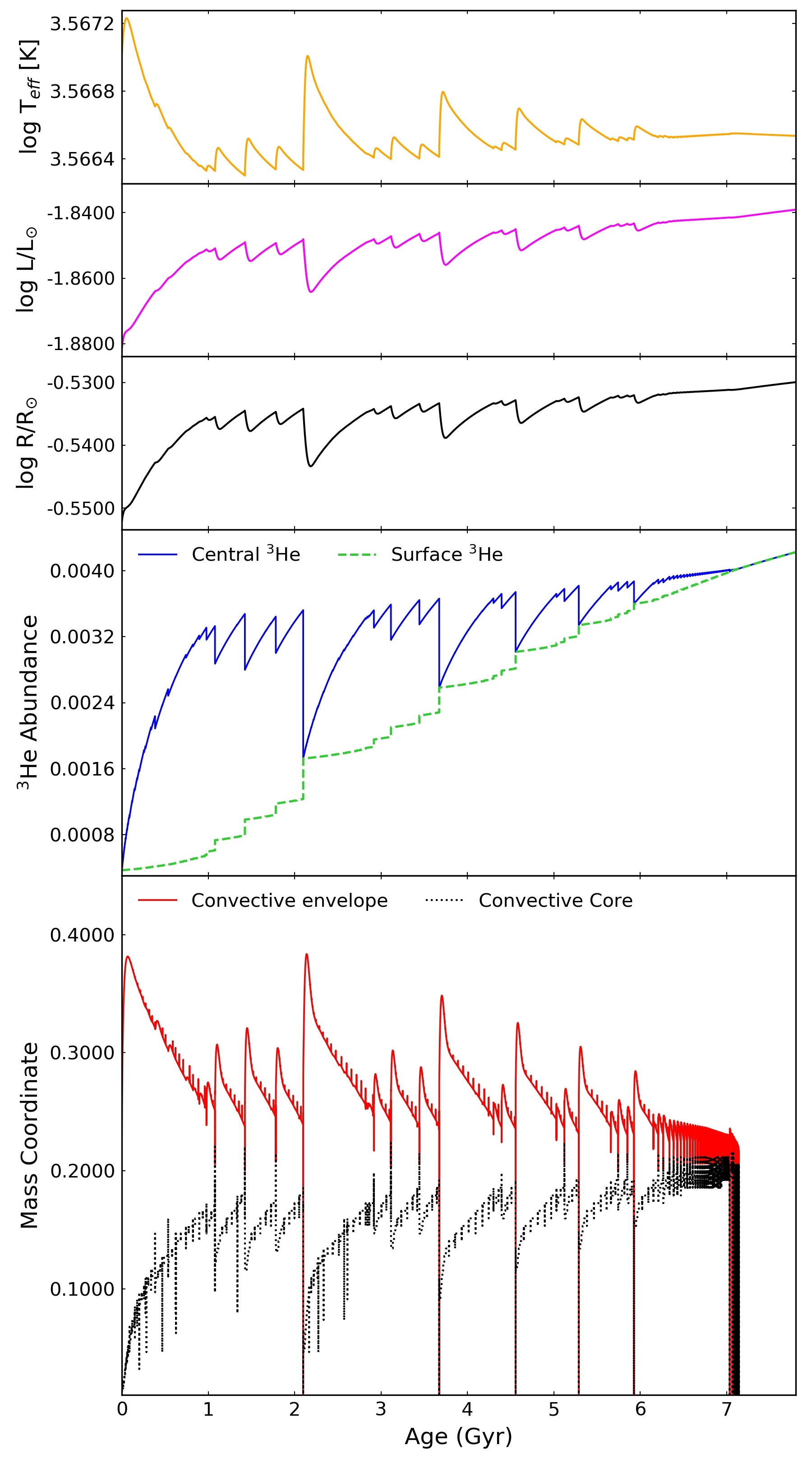}
     
     \vspace*{-6.5pt}
     \caption{HR diagram (top) for the 0.296 M$_{\odot}$, Z = 0.001 model, where each point represents 50,000 years. Also displayed are: the temperature, luminosity, and radius profiles for the same model (middle); and the centre and surface $^3$He abundances (by mass fraction), along with the mass coordinate of the bottom of the convective envelope and the top of the convective core (bottom). A radiative region is found in between. After about $7\,$Gyr, this model has evolved to be fully convective.}
     \label{convmassage}
 \end{figure}

\vspace{8pt}
 The luminosity, effective temperature, and central $^3$He abundance over time are displayed in Fig. \ref{lumtempage} for several models with a metallicity of Z = 0.001. For the sake of clarity, some models are not shown. At this metallicity, the models up to $0.30025\ \textnormal{M}_{\odot}$ undergo fluctuations in luminosity and temperature, some over the course of several billion years, which correlate to the increase and decrease of $^3$He in the core. There is a clear correlation between the dips in central $^3$He abundance and both the peaks in luminosity and the troughs in temperature. The amplitude of the fluctuations generally decreases over time until the models settle into a fully convective state; however, the models also exhibit large changes in luminosity and temperature, corresponding to the larger loops seen in the HR diagrams in Fig. \ref{HR_all}. Figure \ref{convmassage} shows the convective kissing instability for the $0.296\ \textnormal{M}_{\odot},\ \textnormal{Z} = 0.001$ model. The top panel shows the HR diagram, with each point representing 50,000 years, illustrating that the loops occur over a long period of time. The next panels down in the figure give the temperature, luminosity, radius, and $^3$He abundance profiles over time, along with the mass coordinate positions of the top of the convective core and the bottom of the convective envelope; the region in between is radiative. Again, the correlation between these properties is clear. When the bottom of the convective envelope reaches down towards the core, the central $^3$He abundance drops as the material is carried out towards the surface, and correspondingly the surface $^3$He abundance rises. This results in a decrease in luminosity as the $^3$He production is reduced. The temperature rises due to the now fully convective model contracting to compensate for the loss in nuclear energy output, which is represented by the drop in radius. Once the temperature is high enough to resume $^3$He production, the abundance of $^3$He in the core begins to rise along with the luminosity, and the model can increase again in radius. As the radius increases, the surface temperature drops. This repeated process, which results in the fluctuations in temperature and luminosity, produces the loops seen in the HR diagrams in Fig. \ref{HR_all}. After $^3$He reaches its equilibrium abundance, the fluctuations subside and the models remain fully convective. The time period during which the models undergo the convective kissing instability, as well as the amplitude in luminosity and temperature fluctuations, increases with increasing mass. The 1,000 year time step model showed that the smallest pulsations took place on timescales of a few thousand years. 

The models additionally exhibit several events where the bottom of the convective envelope reaches down towards the core but remains disconnected from it. These near-merger events correspond to small drops in central $^3$He abundance and increases in surface $^3$He, and they occur over a period of $\approx$ 1 Myr.

 \begin{table}
    \centering
    \scriptsize
    
    \vspace*{3pt}
        \caption{Model mass ranges for the discontinuity in the luminosity-mass relation.}
        
        \vspace*{2pt}
    \begin{tabular}{lccc}
Age&  Z = 0.01  &   Z = 0.001 & Z = 0.0001\\

[Gyr]& & & \\
 \hline\hline\\
1 & 0.32650 -- 0.33875 M$_{\odot}$ & 0.28600 -- 0.29485 M$_{\odot}$ & 0.28275 -- 0.29000 M$_{\odot}$\\
3&0.33450 -- 0.34325 M$_{\odot}$&0.29275 -- 0.29925 M$_{\odot}$&0.28850 -- 0.29450 M$_{\odot}$\\
5&0.34025 -- 0.34325 M$_{\odot}$&0.29500 -- 0.30125 M$_{\odot}$&0.29050 -- 0.29500 M$_{\odot}$\\
7&0.34175 -- 0.34325 M$_{\odot}$&0.29575 -- 0.30150 M$_{\odot}$&0.29175 -- 0.29450 M$_{\odot}$\\
9&0.34150 -- 0.34350 M$_{\odot}$&0.29775 -- 0.30025 M$_{\odot}$&0.29250 -- 0.29525 M$_{\odot}$\\
 \vspace*{-5pt}
\\ 
\hline
    \end{tabular}
    \label{massranges}
\end{table}

\vspace{3.2pt}
 \subsection{Metallicity dependence}
 
 \vspace{7pt}
 The main effect due to metallicity is the convective kissing instability occurring at larger masses for models with higher metallicity. Correspondingly, the effective temperature is also higher for models with increased metallicity. Figure \ref{0.29} shows the effect of metallicity for one chosen model with mass $m = 0.290\ \textnormal{M}_{\odot}.$ This figure includes the HR diagrams along with the temperature, luminosity, radius, and central $^3$He abundance over time for metallicities of $\ \textnormal{Z} = 0.001$ and $\textnormal{Z} = 0.0001$. A comparison to $\textnormal{Z} = 0.01$ could not be made here as the lowest mass to undergo the convective kissing instability for this metallicity is $0.32650\ \textnormal{M}_{\odot}$. The higher-Z model begins the instability at an earlier age, around $0.3\ \textnormal{Gyr}$, relative to the low-Z model, which begins at around $0.8\ \textnormal{Gyr}$. The fluctuation period subsides and the higher-Z model becomes fully convective by $2.5\ \textnormal{Gyr}$, compared to $5.5\ \textnormal{Gyr}$ for the low-Z model (almost twice as long). The amplitude of the fluctuations found in the luminosity, temperature, radius, and central $^3$He abundance are much higher for the low-Z model.
 
\vspace{4pt}
\subsection{Luminosity-mass relation}

\vspace{8pt}
Figure \ref{lummass} shows the luminosity-mass relation for the three metallicities investigated, at the model ages of 1, 3, 5, 7, and 9 Gyr. A discontinuity is seen at all metallicities. The discontinuity is present over many billions of years and is more prominent at the lower age. The mass ranges in which the discontinuity occurs are given in Table \ref{massranges}. By 9 Gyr, the lower-mass models that underwent the convective kissing instability at 1 Gyr have settled into a fully convective state, and as such the discontinuity shifts to higher masses and higher luminosities over time. Additionally, the discontinuity occurs at higher masses and luminosities for higher metallicities.  Interestingly, the discontinuity is greatly reduced at 7 Gyr for Z = 0.01 to little more than a dip in the luminosity-mass relation, but then appears again at 9 Gyr. Also for Z = 0.01, the highest-mass model to undergo the instability remains 0.34325 M$_{\odot}$ at 3, 5, and 7 Gyr, whereas for the other metallicities the highest mass increases over time.

\vspace{4pt}
\section{Discussion}\label{discussion}

\vspace{8pt} 
The models in this work undergo luminosity, temperature, and radial fluctuations due to the convective kissing instability that was predicted by \textcite{vansaders}, and found in models by \textcite{baraffe} and \textcite{feiden} to be responsible for the Gaia M-dwarf gap. This work finds these fluctuations in the form of loops in the evolutionary tracks in the HR diagrams for models with metallicities down to $\textnormal{Z} = 0.0001$, and our results agree with the conclusion that low-mass stars undergo these periods of fluctuations due to $^3$He production and transport. 

A dip in the luminosity-mass relation was found by \textcite{macdonald}, and by using smaller mass steps, this work finds that there is in fact a discontinuity. We find that the discontinuity occurs at lower masses and luminosities for the lower-metallicity models, as a direct consequence of the convective kissing instability developing at higher masses for higher metallicities. The discontinuity is a result of the merging of the convective envelope and core with the production of $^3$He, and of the model subsequently and periodically switching between partially and fully convective states. This affects the nuclear energy output, which results in the disruption in the luminosity-mass relation as well as the Gaia M-dwarf gap. This discontinuity, which is present over many billions of years, has a great impact on stellar physics as the luminosity-mass relation is no longer differentiable at these masses. For example, in order to constrain the stellar mass function, the slope of the luminosity-mass relation is used (\citealp{pavel}, \citealp{Kroupa+13}). This stellar mass function provides information on the nature of low-mass star formation and the contribution of low-mass stars to the dark matter problem \cite{pavel}. 

\begin{figure}
    \vspace*{1pt}
    \centering
    \includegraphics[width=\columnwidth]{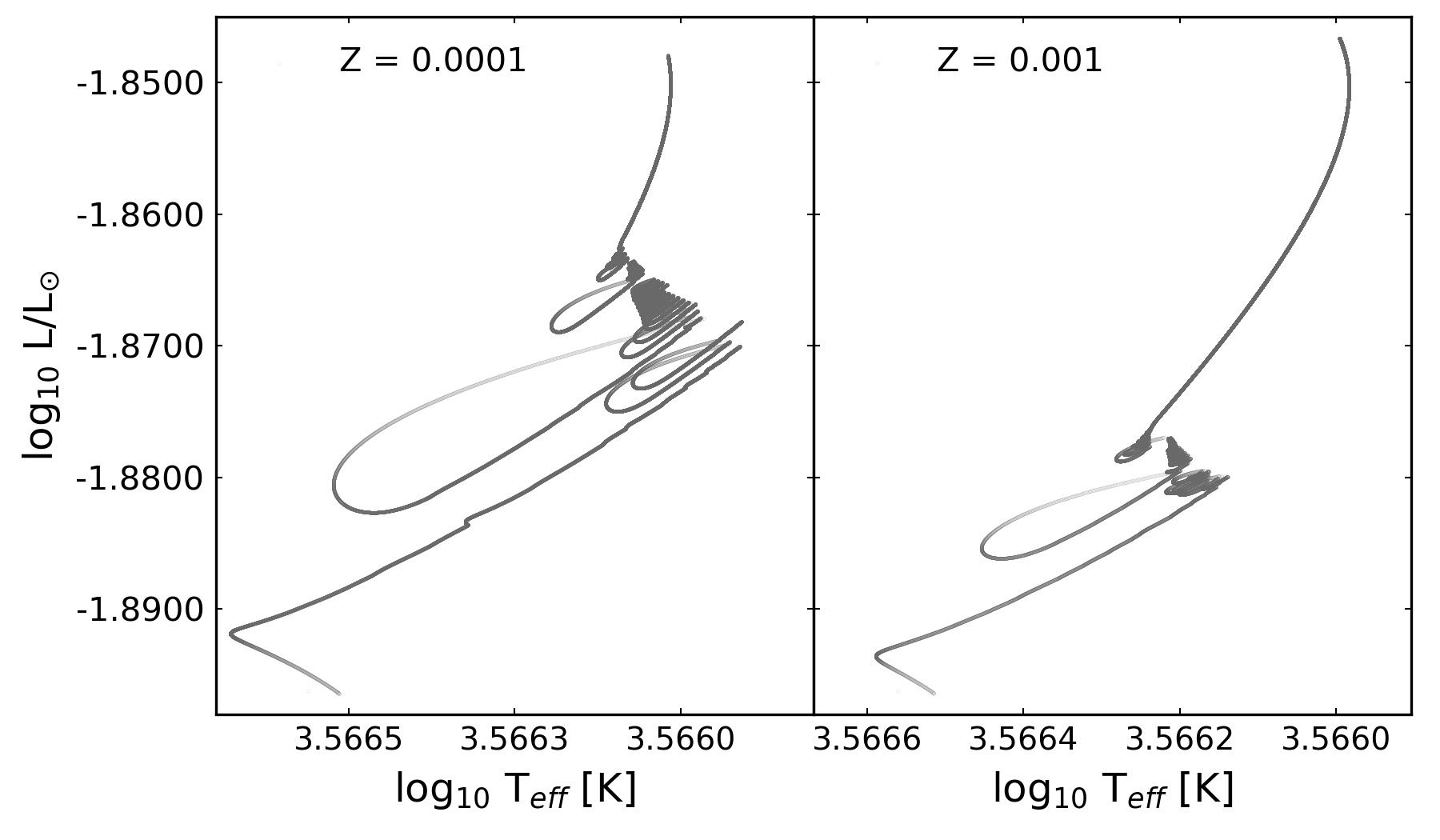}
    \includegraphics[width=\columnwidth]{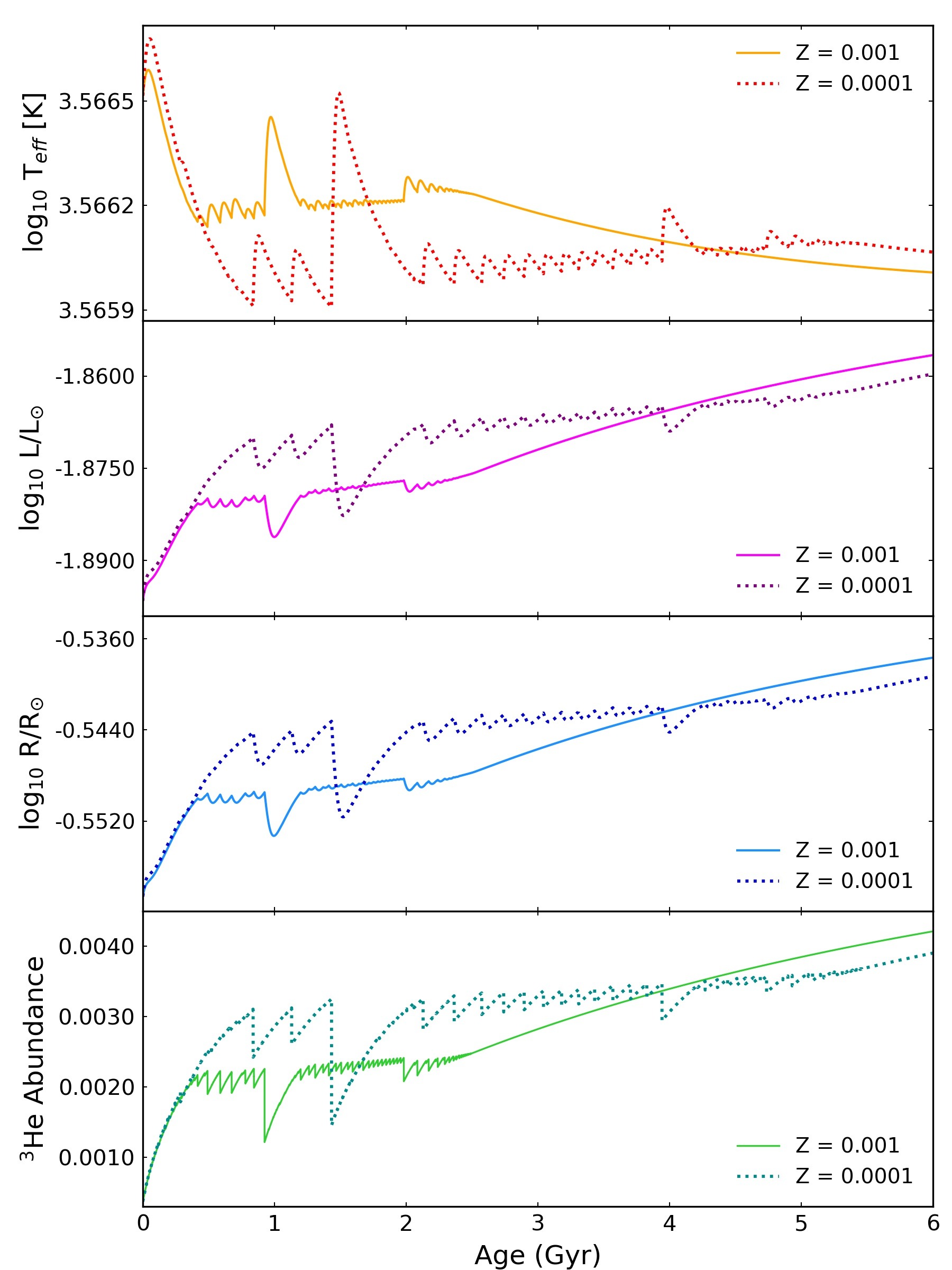}
    
    \vspace*{-0.5pt}
    \caption{HR diagrams for the 0.290 M$_{\odot}$ model with Z = 0.0001 (top left) and Z = 0.001 (top right). Each point represents 50,000 years. Also shown are the temperature, luminosity, radius, and central $^3$He abundance profiles (bottom grid) for the two models.}
    \label{0.29}
\end{figure}

The extra model with a maximum time step of 1,000 years that ran up to 1 Gyr was calculated in order to check if the 50,000 year time step was able to provide an acceptable resolution for the computational time taken. The 1,000 year time step model did reveal that the smallest pulsations took place on timescales of a few thousand years. However, the HR diagram as well as the luminosity and temperature profiles of the 1,000 year model were visually comparable to the 50,000 year time step model. As the 1,000 year model took 82 hours to complete, the computationally quicker 50,000 year time step was adequate for our purposes.

\textcite{feiden} find that the merger events between the convective core and envelope happen four to six times before their models settle into a fully convective state; however, the small time step used in this work reveals a much higher number of merger events. Additionally, some models experience several near-merger events during which the convective envelope reaches down into the radiative layer but not all the way to the core. These mixing events transport $^3$He-rich material into the radiative region, into which the convective envelope reaches and pulls $^3$He outwards to the surface. Whilst the core and envelope do not merge during these events, there exists a million-year-long instability between the two that results in small fluctuations in the $^3$He abundance, as well as in the surface temperature and luminosity. Further investigation is needed to fully resolve and determine the processes occurring during these near-merger events which was beyond the scope of this work.

The convective kissing instability occurs at higher masses for higher metallicities. This was expected due to the increased opacity for higher metallicities. Convection occurs when the radiative temperature gradient is higher than the adiabatic gradient, and by increasing the opacity, a higher temperature (and thus mass) is needed for the growth of a radiative core than for a lower opacity. For lower metallicities, the central temperature required for radiative transport is lower \cite{chabrier1997}, and thus the low-Z models have a radiative core at lower masses than the models with high metallicities. For a given mass, the convective kissing instability occurs earlier in the model lifetime for a higher metallicity since the evolution of the central temperature occurs earlier (see \textcite{chabrier1997}, Fig. 10a). Additionally, the convective kissing instability occurs for longer portions of the model lifetime for lower metallicities (Fig. \ref{0.29}), in agreement with \textcite{feiden} and \textcite{jao}, who note that the observational M-dwarf gap is more prominent at the blue edge of the MS in the HR diagram that represents stars of low metallicities. The instability is found in models with metallicities down to Z = 0.0001, which would be well represented in old globular clusters; this is especially true since the instability period was found to be almost twice as long for the Z = 0.0001 models relative to the Z = 0.001 models. It would be of great importance to find the M-dwarf gap in the observations of these star clusters, although the great distances of the clusters create difficulties in finding these faint low-mass stars. Infrared observations of the nearest globular clusters may be able to achieve this.

\begin{figure*}
    \centering
    \hspace*{-7pt}
    \includegraphics[scale=0.292]{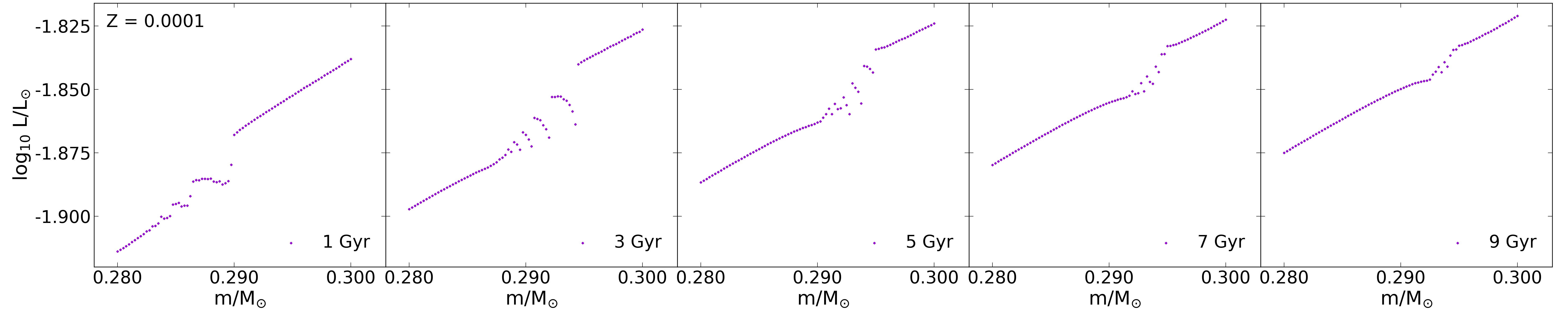}
    
    \hspace*{-7pt}
    \includegraphics[scale=0.292]{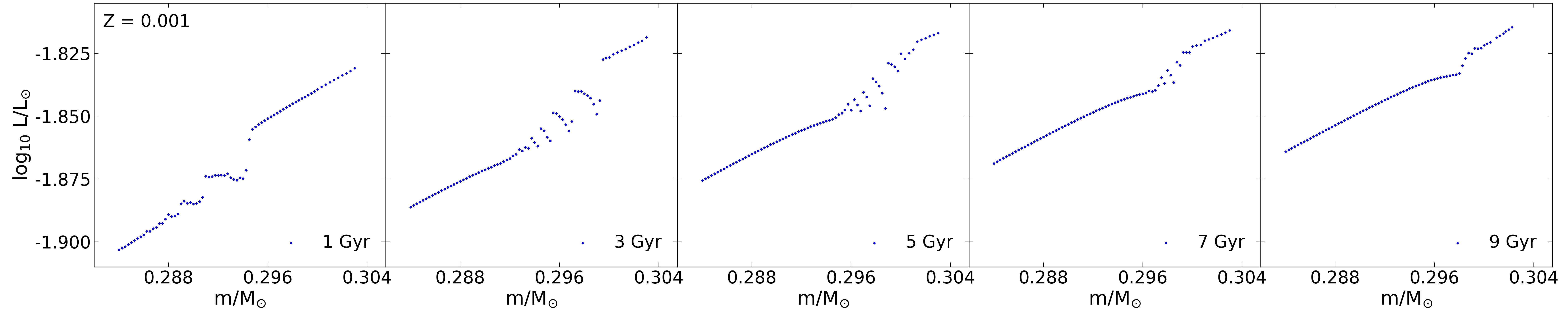}
    
    \hspace*{-7pt}
    \includegraphics[scale=0.292]{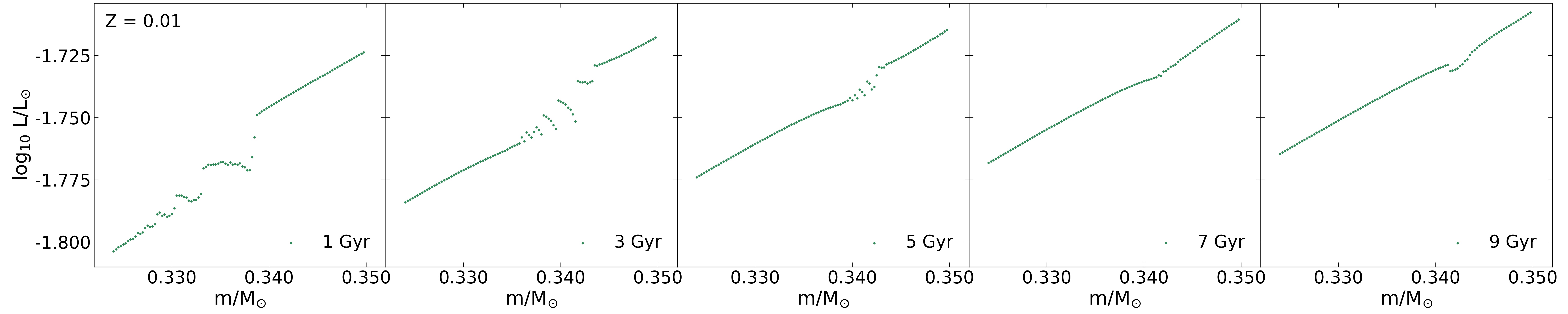}

    \caption{Luminosity-mass relation for metallicities Z = 0.0001 (top), Z = 0.001 (middle), and Z = 0.01 (bottom) at model ages 1, 3, 5, 7, and 9 Gyr. A movie of the full time lapse is available online for the Z = 0.001 model set.}
    \label{lummass}
\end{figure*}

\vspace{5.5pt}
\section{Summary}

\vspace{8.8pt}
Stellar evolution models were computed using three different metallicities over a range of masses in order to reproduce the M-dwarf gap observed in the lower region of the MS. The models used a mass step of $0.00025\ \textnormal{M}_{\odot}$ and a time step of 50,000 years and ran until 10 Gyr. A final model used a 1,000 year time step up to 1 Gyr; however, this was computationally time consuming, and the time step of 50,000 years was adequate for the purposes of this investigation. A discontinuity was found in the luminosity-mass relation for masses $0.32650\ $--$\ 0.34350\ \textnormal{M}_{\odot}$ for the metallicity $\textnormal{Z} = 0.01$, masses $0.28600\ $--$\ 0.30025\ \textnormal{M}_{\odot}$ for $\textnormal{Z} = 0.001$, and masses $0.28275\ $--$\ 0.29525\ \textnormal{M}_{\odot}$ for $\textnormal{Z} = 0.0001$. The HR diagrams for these models show loops during the MS, which results in fluctuations in luminosity and effective temperature. The lower-metallicity models undergo the convective kissing instability at lower masses than the higher-metallicity models, due to the lower opacity and thus the lower temperature (and mass) required for radiative transport in the core. The instability occurs for a longer portion of the model's lifetime for a given mass with decreasing metallicity. 

Comparing the central and surface $^3$He abundances, as well as the mass coordinates of the convective core and envelope, the loops and pulsations correspond to the merging of the core and envelope, resulting in the switching of the model from a partially to fully convective state and the transport of $^3$He from the core to the surface. These loops would lead to the deficiency of stars in the Gaia M-dwarf gap and are due to the production and mixing of $^3$He throughout the star from the merging of the convective core and envelope.

\vspace{10pt}

\begin{acknowledgements}
\\

\\ 
  
We thank the referee for the detailed and thoughtful comments.
\\
PK acknowledges support from the Grant Agency of the Czech Republic under grant number 20-21855S.
\end{acknowledgements}

%-------------------------------------------------------------------

\bibliographystyle{aa}
\nocite{*}
\bibliography{MDwarfGap_ed.bib}

\end{document}